\documentclass[prd,preprintnumbers,twocolumn,amsmath,amssymb,nofootinbib,showpacs]{revtex4}
\usepackage{epsfig,amssymb,amsfonts,verbatim}
\def\bfl{\begin{flushleft}}
\def\efl{\end{flushleft}}
\def\bfr{\begin{flushright}}
\def\efr{\end{flushright}}
\def\bc{\begin{center}}
\def\ec{\end{center}}
\def\be{\begin{equation}}
\def\ee{\end{equation}}
\def\ba{\begin{eqnarray}}
\def\ea{\end{eqnarray}}
\def\baa#1{\begin{array}{#1}}
\def\eaa{\end{array}}
\def\bw{\begin{widetext}}
\def\ew{\end{widetext}}

\def\lb#1{\label{#1}}

\newtheorem{Def}{Def.}[section]

\newtheorem{Prp}[Def]{Proposition}

\begin{document}

%\preprint{arXiv: 0912.xxxx [astro-ph]}

\title{
Comment on
``A limit on the variation of the speed of light arising from quantum
gravity effects''
aka
``Testing Einstein's special relativity with Fermi's short 
hard gamma-ray burst GRB090510''
}

\author{Konstantin G. Zloshchastiev}

\affiliation{National Institute for Theoretical Physics (NITheP)}
%Stellenbosch Institute for Advanced Study,
%Stellenbosch, South Africa}
\affiliation{Institute of Theoretical Physics, University of Stellenbosch,
Stellenbosch 7600, South Africa}

%\affiliation{Department of Physics, National University of Singapore,
%Singapore 117542 %, Republic of Singapore
%}

%\affiliation{Department of Theoretical Physics, Dnepropetrovsk National University, Dnepropetrovsk 49050, Ukraine}

%\date{~ ~~~~~~~~~~~~~~~~~~~~~~}
%\date{Submitted on 10 Mar 2009}
%\date{~Received: 26 May 2000 [PRL], 1 June 2000 [LANL] ~}
%\date{Received \today}

%\scriptsize%\footnotesize

\begin{abstract}
Recently the Fermi GBM and LAT Collaborations
reported  their new observational data
disfavoring quite a number of the quantum gravity theories,
including the one suggesting the nonlinear (logarithmic) modification
of a quantum wave equation.
We show that the latter is still far from being ruled out:
it is not only able to explain the new data but also its phenomenological implications 
turn out to be more vast (and more interesting) than one
%, including myself, 
expected before.
\end{abstract}

\pacs{04.60.Bc, 98.70.Sa}
\maketitle

%\narrowtext

\large
%\newpage

One of the outcomes of 
theory 
%proposed in %the early versions (v2-) of Ref. 
\cite{Zloshchastiev:2009zw}
%which was derived 
is that under some conditions
the LIV corrections
to propagation speed of a particle are given by
\be\lb{e-disp1}
v/c 
=
1 \pm \frac{E}{E_{\text{QG}}} + {\cal O} (E^2/E_{\text{QG}}^2),
%\qquad \xi \equiv \pm 1, 
\ee
which, after integrating
$\partial E/\partial p = v (E)$,
leads to the vacuum dispersion relation
of the form
\be\lb{e-ddr}
\left|
\frac{p^2 c^2}{ E^2}
-1
\right|
=
\frac{E}{E_{\text{QG}}}
+ {\cal O} (E^2/E_{\text{QG}}^2)
%\right]
,
\ee
and the agreement with the existing observational
data (at that time) was established.

In  more recent Ref. \cite{Collaborations:2009zq}
 it was suggested
to rule out such dispersions for 
$E_{\text{QG}} = M_\text{Planck} c^2$ 
on experimental grounds
- essentially,
because the predicted absolute value of the series coefficient 
for a linear term, unity, seems to be  significantly larger
than the one suggested by observations.

After that, more thorough analysis of our theory's predictions
has been done.
%, starting from the third version of Ref. \cite{Zloshchastiev:2009zw}.The essential conclusion is 
It turns out that
for the extremely 
ultrarelativistic particles, such as the high-energy photons 
from the not-very-distant GRB's,
the above-mentioned dispersion relations are unlikely 
applicable because
for that physical situation
our theory suggests different ones.
Let us see it.
To begin, the {\it primary} outcome of our theory is not the dispersion relations
themselves
but the expression for the invariant,
%\be\lb{e-inv}
$
d\tau^2/(E - E_0)^2
%= \frac{\beta_1}{\beta_2} = \frac{d\tau_2}{E_2 - E_0}= \text{inv}
,
$
%at least in the leading order w.r.t. $E/E_0$,
where $\tau$ is the proper time, 
$E_0$ is the energy of the vacuum of a theory, in our case
$E_0 = \pm E_{\text{QG}}$.
From this one  concludes that for any two particles
\be\lb{e-b12}
\frac{d\tau_2}{d\tau_1}
=
%\frac{\beta_1}{\beta_2} = 
\frac{E_2 - E_0}{E_1 - E_0}
,
\ee
where $\tau_i$ and $E_i$ are the proper time and energy for the $i$th particle.
When neglecting the cosmological effects it simplifies to
\be\lb{e-vbra}
\frac{v_1}{v_2}
\sqrt{
\frac{c^2 - v_2^2}{c^2 - v_1^2}
     }
= \frac{E_2 - E_0}{E_1 - E_0}
,
\ee
where $v_i = d x_i/ d t_i$, by $t$ we denote the time coordinate measured
by a distant observer.

%In what follows we will be interested in their ratio $v_1/v_2$ as a function of $v_2$ and energies (which will be later used for deriving the dispersion relations).
This equation reveals the following subtlety: if 
%for our future purposes we assume that 
the particles are essentially relativistic 
or even ultrarelativistic and their velocities are nearly the same,
then the value of a square root in the equation
above crucially depends on whether the ratio $v_1/c$ tends to unity ``stronger'' than
$v_1/v_2$. Thus, there exist  two limit regimes:
{\it linear} or standard relativistic - when the ratio
$v_1/v_2$ approaches one ``stronger'' than $v_1/c$ does, 
and {\it non-perturbative} or extreme ultrarelativistic - when it is other way around.
In the former case one can approximate
the square root in Eq. (\ref{e-vbra}) by unity, then eventually one arrives at the 
dispersions (\ref{e-disp1}) and (\ref{e-ddr}).
In the latter regime such approximation is not valid, instead, one should perform
the non-perturbative calculation to
eventually obtain:
\be\lb{e-vnpt-s}
\frac{v^{(s)} }{c^{(s)}}=
%\left[
1
+
\frac{\chi^2-1}{\chi^2}
\frac{E}{E_0}
%+  \frac{(\chi^2-1)(\chi^2 - \frac{3}{2})}{\chi^4} \frac{E^2}{E_0^2}
+
{\cal O} (E^2/E_{\text{QG}}^2)
%\right]
,
\ee
where $c^{(s)} \equiv c/\chi$ is the ``renormalized'' speed of light,
$\chi$ is the emerging parameter representing
the  ratio of the ``bare'' speed of light and the ``renormalized'' one.
Without specifying a concrete model of quantum gravity, 
our theory can not provide the exact value of  $\chi$, it gives only the 
%lower bound and 
approximate range for subluminal particles: $\chi^2 \gtrapprox 1$.
This dispersion mode is not ruled out by  
the Fermi's  data - those can only put further bounds for 
the parameter. 
More details as well as some new predictions can be found in the revised version 
of Ref. \cite{Zloshchastiev:2009zw} (v3+).

This work was supported under a grant of the National Research Foundation of South Africa. 

\def\AnP{Ann. Phys.}
\def\APP{Acta Phys. Polon.}
\def\CJP{Czech. J. Phys.}
\def\CMPh{Commun. Math. Phys.}
\def\CQG {Class. Quantum Grav.}
\def\EPL  {Europhys. Lett.}
\def\IJMP  {Int. J. Mod. Phys.}
\def\JMP{J. Math. Phys.}
\def\JPh{J. Phys.}
\def\FP{Fortschr. Phys.}
\def\GRG {Gen. Relativ. Gravit.}
\def\GC {Gravit. Cosmol.}
\def\LMPh {Lett. Math. Phys.}
\def\MPL  {Mod. Phys. Lett.}
\def\Nat {Nature}
\def\NCim {Nuovo Cimento}
\def\NPh  {Nucl. Phys.}
\def\PhE  {Phys.Essays}
\def\PhL  {Phys. Lett.}
\def\PhR  {Phys. Rev.}
\def\PhRL {Phys. Rev. Lett.}
\def\PhRp {Phys. Rept.}
\def\RMP  {Rev. Mod. Phys.}
\def\TMF {Teor. Mat. Fiz.}
\def\prp {report}
\def\Prp {Report}

\def\jn#1#2#3#4#5{{#1}{#2} {\bf #3}, {#4} {(#5)}} %PRD
%\def\jn#1#2#3#4#5{{#1}{#2} {#3} {(#5)} {#4}}   %PLB style
% #1 tittle  #2 ser  #3 vol  #4 page  #5 year

\def\boo#1#2#3#4#5{{\it #1} ({#2}, {#3}, {#4}){#5}}
%\def\boo#1#2#3#4#5{ #1 ({#2}, {#3}, {#4}){#5}}  %PLB style
% #1 tittle  #2 publisher  #3 place  #4 year  #5 page/, p.789/

%\def\jn#1#2#3#4#5{{#1}{#2} {\bf #3}, {#4} {(#5)}}
% #1 tittle  #2 ser  #3 vol  #4 page  #5 year
%\def\boo#1#2#3#4#5{{\it #1} ({#2}, {#3}, {#4}){#5}}
% #1 tittle  #2 publisher  #3 place  #4 year  #5 page/, p.789/

%\newpage


\begin{thebibliography}{99}
%\footnotesize

%\cite{Zloshchastiev:2009zw}
\bibitem{Zloshchastiev:2009zw}
  K.~G.~Zloshchastiev,
 ``Logarithmic nonlinearity in generally covariant quantum theories: Origin of time and observational consequences,''
  arXiv:0906.4282 [hep-th].
  %%CITATION = ARXIV:0906.4282;%%
  
  %\cite{Collaborations:2009zq}
\bibitem{Collaborations:2009zq}
A.~A.~Abdo {\it et al.}  [Fermi LAT/GBM Collaborations],
``A limit on the variation of the speed of light arising from quantum
gravity effects,''
Nature {\bf 462} (2009) 331-334;
%a.k.a.
A.~A.~Abdo {\it et al.},
%  [Fermi LAT/GBM Collaborations],
``Testing Einstein's special relativity with Fermi's short hard gamma-ray burst GRB090510,''
  arXiv:0908.1832 [astro-ph.HE].
  %%CITATION = ARXIV:0908.1832;%%

\end{thebibliography}
\end{document}